\documentclass[twocolumn,preprintnumbers,amsmath,amssymb,superscriptaddress]{revtex4-2}
\UseRawInputEncoding

\usepackage{latexsym,amssymb,amsthm,amsmath,epsfig}
\usepackage[left=2.00cm, right=2.00cm, top=2.00cm, bottom=2.00cm]{geometry}
\usepackage{float}
\usepackage{hyperref}
\hypersetup{
    citecolor=red,
    colorlinks=true,
    linkcolor=blue,
    filecolor=blue,      
    urlcolor=blue,
}

\begin{document}

\title{Magnomechanically controlled Goos-H\"{a}nchen shift in cavity QED}

\author{Muhammad Waseem}
\affiliation{Department of Physics and Applied Mathematics, PIEAS, Nilore, Islamabad 45650, Pakistan}
\affiliation{Center for Mathematical Sciences, PIEAS, Nilore, Islamabad 45650, Pakistan}

\author{Muhammad Irfan}
\email[Corresponding author: ]{m.irfanphy@gmail.com}
\affiliation{Department of Physics and Applied Mathematics, PIEAS, Nilore, Islamabad 45650, Pakistan}
\affiliation{Center for Mathematical Sciences, PIEAS, Nilore, Islamabad 45650, Pakistan}

\author{Shahid Qamar}
\affiliation{Department of Physics and Applied Mathematics, PIEAS, Nilore, Islamabad 45650, Pakistan}
\affiliation{Center for Mathematical Sciences, PIEAS, Nilore, Islamabad 45650, Pakistan}


\date{\today }

\begin{abstract}
Phenomena involving interactions among magnons, phonons, and photons in cavity magnomechanical systems have attracted considerable attention recently, owing to their potential applications in the microwave frequency range.
One such important effect is the response of a probe field to such tripartite interaction between photon-magnon-phonon.
In this paper, we study Goos-H\"{a}nchen shift (GHS) of a reflected probe field in a cavity magnomechanical system.
We consider a YIG sphere positioned within a microwave cavity.
A microwave control field directly drives the magnon mode in YIG sphere, whereas the cavity is driven via a weak probe field. 
Our results show that the GHS can be coherently controlled through magnon-phonon coupling via the control field.
For instance, GHS can be tuned from positive to negative by tuning the magnon-phonon coupling.
Similarly, the effective cavity detuning is another important controlling parameter for GHS.
Furthermore, we observe that the enhancement of GHS occurs when magnon-phonon coupling is weak at resonance, and when the magnon-photon coupling is approximately equal to the loss of microwave photons.
Our findings may have potential significance in applications related to microwave switching and sensing.

\end{abstract}

\maketitle

\section{introduction}
The cavity magnomechanical system consisting of magnons in a single-crystal yttrium iron garnet (YIG) sphere strongly coupled to the cavity mode has been theoretically proposed and experimentally demonstrated ~\cite{zhang_cavity_2016}.
Such a system has emerged as an important frontier in the realm of cavity quantum electrodynamics (QED), drawing substantial attention in recent times~\cite{zhang_cavity_2015,goryachev_high-cooperativity_2014, li_hybrid_2020,soykal_strong_2010,bai_spin_2015,zhang_strongly_2014}.
It offers a unique platform for exploring interactions among photon, magnon, and phonon modes.
Such interactions led to some interesting outcomes such as the generation of entanglement~\cite{li_magnon-photon-phonon_2018, hussain_entanglement_2022}, preparation of squeezed states~\cite{li_squeezed_2019,li_squeezing_2023}, coherent superposition and bell states~\cite{yuan_steady_2020}. 

Meanwhile, studies on the response of microwave field to the system arising from the coupling of the magnon, phonon, and the cavity microwave photon, reveal the magnon-induced absorption, as well as the magnomechanically induced transparency~\cite{li_phase_2020,kong_magnetically_2019,new_wang_magnon-induced_2018,liu_room-temperature_2019,wen_tunable_2019, lu_exceptional-point-engineered_2021,noauthor_kerr1,ullah_tunable_2020}.
These phenomena originate from the internal constructive and destructive interference that can be interpreted by analogy with the optomechanical induced absorption and transparency, respectively, in the cavity optomechanics~\cite{hou_optomechanically_2015,zhang_optomechanically_2017, agarwal_electromagnetically_2010}.
Tunable slow and fast light has also been demonstrated in the cavity magnomechanics~\cite{kong_magnetically_2019, new_wang_magnon-induced_2018, liu_room-temperature_2019,wen_tunable_2019}.  
In this paper, we focus on another crucial aspect of optical response in a cavity magnomechanical system, known as the Goos-H\"{a}nchen shift (GHS). 

In classical optics, the Goos-H\"{a}nchen shift occurs when a classical electromagnetic light beam reflects from the interface of two optically different media. 
It is a lateral shift of reflected light beam from the actual point of reflection at the interface and first reported in an experiment by Goos and H\"{a}nchen~\cite{goos_neuer_1947,picht_beitrag_1929}.
The GHS has magnificent applications in optical switching~\cite{wang_all-optically_2013}, optical sensors~\cite{hsue_lateral_1985,wang_electric_2008}, beam splitters~\cite{wang_all-optically_2013}, optical temperature sensing~\cite{chen_optical_2007}, acoustics~\cite{declercq_double-sided_2004}, seismology~\cite{wang_influence_2015}, and theory of waveguides~\cite{white_effect_1977}.
The GHS can be positive or negative depending on the properties of the system under consideration. 
So far various quantum systems have been investigated for the manipulation of GHS, such as atom-cavity QED~\cite{ziauddin_coherent_2010,ziauddin_gain-assisted_2011,wang_control_2008}, quantum dots~\cite{idrees_enhancement_2023,darkhosh_incoherent_2022}, cavity optomechanics~\cite{ullah_flexible_2019,khan_investigation_2020,ghaisuddin_enhancement_2021}, two-dimensional quantum materials~\cite{shah_electrically_2022,shah_quantization_2022,shah_quantized_2021}, and a spin-polarized neutron reflecting from a film of magnetized material~\cite{de_haan_observation_2010}. 
However, to the best of our knowledge, manipulation of GHS in a cavity magnomechanical system has not been reported yet which could have possible potential applications in sensing and switching.

In this paper, we investigate the manipulation of GHS in the reflected portion of the incident probe field in a cavity magnomechanical system using the stationary phase method. 
We show that the magnon-phonon coupling strength, controlled via an external microwave driving field, flexibly alters the GHS from negative to positive.
The negative GHS becomes larger at weak magnon-phonon interaction strength.
We also show that GHS can be controlled by tuning the effective cavity detuning.
By varying the effective detuning at zero magnon-phonon coupling strength, GHS can be effectively tuned from positive to negative.
Finally, we explore the effects in both weak and strong coupling limits, determined by the ratio of magnon-photon coupling to the lifetime of microwave photons.
We find that the optimum value to achieve enhanced GHS corresponds to magnon-photon coupling of approximately the same strength as the cavity decay rate.  

The rest of the paper is organized as follows: In section II, we explained the physical system. Section III deals with results and discussion. Finally, we conclude in section IV.


 \section{system model and Hamiltonian}
\begin{figure}[t]
\begin{tabular}{@{}cccc@{}}
\includegraphics[width=3.25 in]{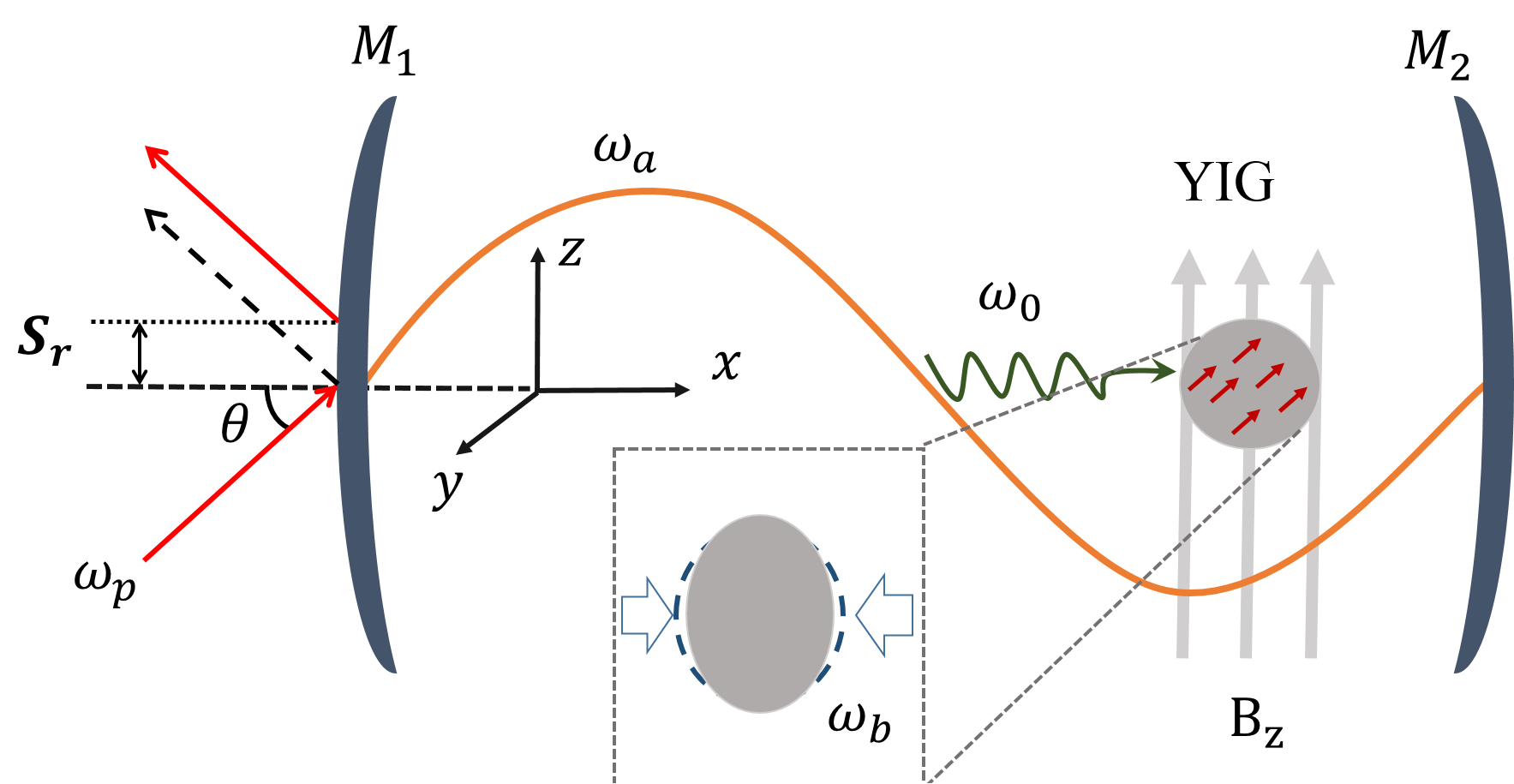}
\end{tabular}
\caption{Schematic diagram of the physical system. 
A YIG sphere is positioned inside a single-mode microwave cavity, near the maximum magnetic field of the cavity mode, and simultaneously subjected to a uniform biased magnetic field $B_{z}$. This arrangement establishes the magnon-photon coupling.
A microwave field of frequency $\omega_0$ is applied along the $x$-direction by an external drive magnetic field $B_x$ to enhance the magnon-phonon coupling. The magnetic field of the cavity mode $B_y$, biased magnetic field $B_{z}$, and drive magnetic field $B_x$ are mutually perpendicular at the site of the YIG sphere.  
The incident probe field falls along $x$-direction on the wall of the mirror $M_1$ at angle $\theta$, which is reflected with positive or negative GHS denoted by $S_{r}$.} 
\label{fig:1}
\end{figure}

We consider a cavity-magnomechanical system that consists of a single-mode cavity of frequency $\omega_a$ with a YIG sphere placed inside the cavity as shown in Fig.~\ref{fig:1}.
Both nonmagnetic mirrors $M_1$ and $M_2$ are kept fixed, assuming $M_2$ is perfectly reflecting while $M_1$ is partially reflecting.
Both mirrors have thickness $d_1$ and permittivity $\epsilon_1$.
The effective cavity length is $d_2$ and effective cavity permittivity in the presence of a YIG sphere is $\epsilon_2$.
Therefore, our system is effectively a three-layer structure comprising two mirrors and an intracavity medium similar to atomic system~\cite{wang_control_2008} and cavity optomechanics~\cite{Muhib_2019}.
In Fig.~\ref{fig:1}, a uniform bias magnetic field along the $z$-direction is applied on the YIG sphere, which excites the magnon modes of frequency $\omega_m$.
These magnon modes are coupled with the cavity field through magnetic dipole interaction.
The excitation of the magnon modes inside the sphere varies the magnetization, resulting in the deformation of its lattice structure.
This magnetostrictive force causes vibrations of the YIG sphere with phonon frequency $\omega_b$, which establishes magnon-phonon interaction.

Usually, single-magnon magnomechanical coupling strength is very weak~\cite{wang_bistability_2018}.
However, we consider that the magnon mode of the YIG sphere is directly driven by a strong external microwave source having frequency $\omega_0$ and amplitude $E_{d}=\sqrt{5 N} \gamma B_{0}/4$. 
Here, $\gamma$ is the gyromagnetic ratio, $N$ is the total number of spin inside the YIG sphere, and $B_{0}$ is the magnitude of the external drive field along $x$-direction. 
This microwave drive plays the role of a control field in our model and enhances the magnomechanical (magnon-phonon) interaction of the YIG sphere.
Additionally, the cavity is probed by a weak field with frequency $\omega_p$, incident from vacuum $\epsilon_0$ at an angle $\theta$ along the $x$-axis.
The amplitude of the probe field is $E_p=\sqrt{2 P \kappa_a/( \hbar \omega_p)}$.
Here, $P$ is the power of the probe field, and $\kappa_a$ is the cavity decay rate.
The probe light is reflected back from the surface of the mirror $M_1$ with some lateral displacement along the $z$-axis known as GHS and denoted by $S_r$.


To investigate GHS, we employ stationary phase theory, in which a well-collimated probe field with sufficiently large linewidth can be considered as a plane wave.
Under the stationary phase theory, the GHS in the reflected probe laser beam is given by~\cite{artmann_berechnung_1948,li_negative_2003}:
\begin{align}\label{E1}
 S_r=&-\frac{\lambda_p}{2 \pi} \frac{d\phi_r}{d \theta},
\end{align}
where, $\lambda_p$ is the wavelength of incident probe field, $\phi_r$ is the phase of the TE polarized reflection coefficient $R(k_z, \omega_p)$. Here $k_{z}= k \sin{\theta}$ with $k=2 \pi / \lambda_p$.  
Equation~(\ref{E1}) can be expressed in a more explicit form such that~\cite{wang_large_2005}
\begin{align}\label{E2}
 S_{r}=& -\frac{\lambda_{p}}{2 \pi |R|^2}  [ \text{Re}(R) \frac{d }{d \theta} \text{Im} (R) + \text{Im} (R) \frac{d}{d \theta} \text{Re}(R)].
 \end{align}
The reflection coefficient $R=R(k_z, \omega_p)$ used in the above equation can be derived using standard transfer matrix theory for three-layer structure~\cite{wang_control_2008} and is given by
\begin{align}\label{E3}
 R=& \frac{q_{0} (Q_{22} -Q_{11}) -(q_{0}^{2} Q_{12}-Q_{21}) }{q_{0} (Q_{22} + Q_{11}) -(q_{0}^{2} Q_{12}+Q_{21})}.
 \end{align}
This transfer matrix approach is well established in the context of GHS in atomic systems~\cite{wang_control_2008, ziauddin_coherent_2010}, cavity optomechanics~\cite{Muhib_2019, khan_investigation_2020}, quantum dots~\cite{idrees_enhancement_2023}, among others.
Here $q_{0}=\sqrt{\epsilon_{0} -\sin^{2} \theta}$ and $Q_{ij}$ ($i,j=1,2$) are the elements of total transfer matrix:
\begin{align}\label{E4}
 Q(k_{z}, \omega_p)=& m_{1} (k_{z}, \omega_{p},d_{1}) m_{2} (k_{z}, \omega_{p},d_{2}) m_{1} (k_{z}, \omega_{p},d_{1}),
 \end{align}
Here, $d_1$ is the thickness of mirror $M_1$ and $M_2$ while $d_2$ is the effective cavity length containing the YIG sphere. The element $m_j$ relates the input and output of the electric field associated with the probe field propagating through the cavity and is given by:
\begin{align}\label{E5}
 m_j(k_{z}, \omega_{p},d_{1}) = & \begin{bmatrix}
            \cos(k_{j}^{x} d_{j}) & i \sin(k_{j}^{x} d_{j}) k / k_{j}^{x}   \\
            i \sin(k_{j}^{x} d_{j}) k_{j}^{x}/k & \cos(k_{j}^{x} d_{j}) 
        \end{bmatrix},
 \end{align}
where $k_{j}^{x}=( \omega_p / c ) \sqrt{\epsilon_{j}-\sin^{2} \theta}$ is the $x$-component of the wave number of probe field. 
Here, we use the definition of $k=\omega_p / c$ with $c$ being the speed of light in vacuum.
Whereas $\epsilon_j$ represents the susceptibility of the $j^{th}$ layer of the medium. 
The effective permittivity of the cavity is determined by the non-linear susceptibility $\chi$ as $\epsilon_{2}=1 + \chi$.
The susceptibility depends on the non-linear interaction between the cavity field, magnon, and phonons in the presence of the microwave drive.
As a result, the resonance conditions for the probe field are modified, resulting in a controllable absorption and dispersion.
Therefore, the reflection properties of the probe field strongly depend on the cavity magnomechanical interaction.

Next, we calculate $\chi$ for the cavity magnomechanical system to study the reflection properties of the probe field.
We consider the dimensions of the YIG sphere much smaller than the wavelength of the microwave so that the influence of radiation pressure in the system can be negligible. 
In a frame rotating with the driving frequency $\omega_0$, the total Hamiltonian (in unit of $\hbar=1$) of the system under rotating-wave approximation becomes:
 \begin{align}\label{E6}
 H=&\Delta_a a^{\dag}a+\Delta_m m^{\dag}m+\omega_b b^{\dag}b \nonumber\\
 &+g_{ma}(a^{\dag}m+am^{\dag})
 +g_{mb} m^{\dag}m (b+b^{\dag}) \nonumber\\
 &+i(E_{d} m^{\dag}+E_{p}e^{-i \delta t} a^{\dag}-H.c.).
\end{align}
Here, $a(a^{\dag})$, $m(m^{\dag})$, and $b(b^{\dag})$ are the annihilation(creation) operators of the cavity mode, the magnon mode, and the mechanical mode, respectively.
The detuning of the cavity field, magnon mode, and probe field from the control field is $\Delta_{a}=\omega_{a}-\omega_{0}$, $\Delta_{m}=\omega_{m}-\omega_{0}$, and  $\delta=\omega_{p}-\omega_{0}$, respectively.
The magnomechanical coupling rate $g_{mb}$ characterizes the interaction between the magnon and phonon modes, whereas $g_{ma}$ determines the photon-magnon coupling strength. 
\begin{figure}[t]
\begin{tabular}{@{}cccc@{}}
\includegraphics[width=3.25 in]{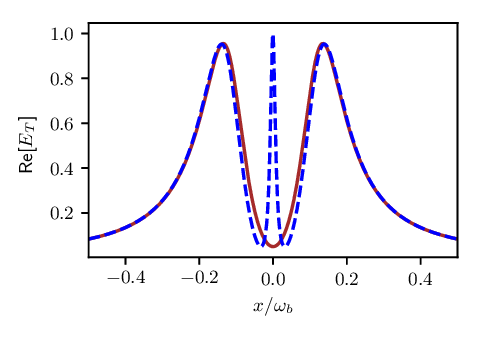}
\end{tabular}
\caption{Absorption spectrum of the probe field as a function of effective detuning $x=\delta-\omega_b$. Here, the solid curve is at magnon-phonon coupling $G_{mb}=0$ while the dashed curve is at $G_{mb}=2 \pi \times 0.5$MHz. All other parameters are the same as given in the text.} 
\label{fig:chi}
\end{figure}

To understand the dynamics of the system, we write, within the semi-classical limit, the Heisenberg-Langevin equations:  
\begin{align}\label{E7}
\dot{a}=&-(i\Delta_a+\kappa_a)a-ig_{ma}m + E_{p} e^{-i \delta t} ,\nonumber\\
\dot{m}=&-(i\Delta_s+\kappa_m)m-i g_{ma}a+E_{d},\nonumber\\
\dot{b}=&-(i\omega_b + \gamma_b)b -i g_{mb} m^{\dag}m,
\end{align}
where 
\begin{align}
\Delta_s=&\Delta_m +g_{mb} (b_s+b_s^{*}),\nonumber
\end{align}
is the effective magnon-phonon detuning.
In the above equations, we take into account the decay of the cavity mode $\kappa_a$, the dissipation of magnon mode $\kappa_m$, and the dissipation of mechanical mode $\gamma_b$.
Since we are interested in studying the mean response of this system to the applied probe field, we have neglected the quantum input noise and thermal noise.
Using a semiclassical perturbation framework, we consider that the probe microwave field is much weaker than the control microwave field.
As a result, we expand each operator as the sum of its steady state value $o_s$ and a small fluctuation $\delta o(t)$, where $o=(a,b,m)$.
Then steady-state values of the dynamical variables become:

\begin{align}
\label{steady}
a_s=&\frac{-i g_{ma} m_{s}}{i \Delta_a + \kappa_a},\nonumber\\
m_s=&\frac{-i g_{ma} a_{s} + E_{d}}{i \Delta_s + \kappa_m},\nonumber \\
b_s=&\frac{-i g_{mb} |m_{s}|^2}{i \omega_b + \gamma_b}.
\end{align}
Considering the perturbation induced by the input probe field up to the first-order term and eliminating the steady state values, we obtain the linearized equations of motion:

\begin{align}\label{lin}
\delta \dot{a}=&- \kappa_a \delta a-ig_{ma} \delta m + E_{p} e^{-i x t} ,\nonumber\\
\delta \dot{m}=&- \kappa_m \delta  m-i g_{ma} \delta a -i g_{mb} m_{s} \delta b , \nonumber\\
\delta \dot{b}=&- \gamma_b \delta b -i g_{mb} m_{s}^{*} \delta m,
\end{align}
where $x=\delta-\omega_b$ is the effective detuning. While driving the above equation, we introduce the slowly varying operator for the linear terms of fluctuation as $\delta a = \delta a e^{-i \Delta_{a} t}$, $\delta m = \delta m e^{-i \Delta_{s} t}$, and $\delta b = \delta b e^{-i \omega_{b} t}$.
We also consider that the microwave field driving the magnon is at the red sideband ($\omega_b \approx \Delta_a \approx \Delta_s$) under rotating-wave approximation, which actually leads to optimal cooling~\cite{agarwal_electromagnetically_2010}.

In order to solve Eq.~(\ref{lin}), we apply an ansatz $\delta o = o_1 e^{-i x t} + o_2 e^{i x t}$ with $o=(a, m, b)$. 
As a result, we obtain the amplitude $a_1$ of the first-order sideband of the cavity magnomechanical system for a weak probe field~\cite{note2}:
\begin{align}\label{a1}
a_1=&\frac{E_p}{(\kappa_a - i x) + \frac{g_{ma}^{2} (\gamma_b - i x) }{(\gamma_b - i x) (\kappa_m - i x) + G_{mb}^2}}.
\end{align}
Here, $G_{mb}=g_{mb} m_s$ is the effective magnomechanical coupling coefficient, which can be tuned by an external magnetic field at fixed $g_{mb}$. 
Furthermore, it is not necessary to consider the expression of $a_2$ as it pertains to four-wave mixing with frequency $\omega_p - 2 \omega_0$ for the driving field and the weak probe field. 
Then, using the input-output relation, we obtain $E_{T}=E_{in}-\kappa a_{1}$~\cite{gardiner_input_1985}.
The output field is related to the optical susceptibility as $\chi=E_{T}=\kappa_{a} a_{1}/ E_{p}$~\cite{khan_investigation_2020, ghaisuddin_enhancement_2021,chen_electromagnetically_2023, agarwal_electromagnetically_2010, li_transparency_2016}.
Here, $\chi$ is a complex expression that defines the quadrature of the field $E_{T}$ containing real and imaginary parts. The quadrature is defined as $\chi = \chi_{r}+i\chi_{i}$ and can be measured by homodyne techniques. The real term displays the absorption spectrum, while the imaginary term displays the dispersion spectrum of the probe field.
\begin{figure}[t]
\begin{tabular}{@{}cccc@{}}
\includegraphics[width=3.25 in]{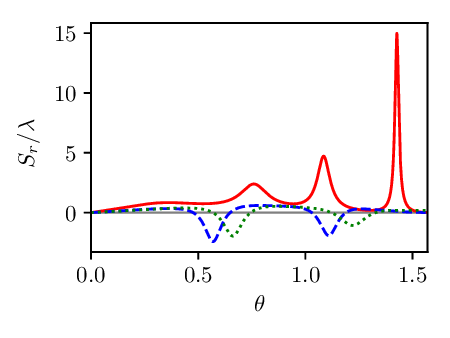}
\end{tabular}
\caption{Normalized GHS $S_{r}/\lambda$ as a function of incident angle at resonance condition $x=0$: Solid curve shows that GHS is always positive when $G_{mb}=0$. Whereas dashed and dotted curves represents the results at $G_{mb}=2\pi \times 0.05 $MHz, and $G_{mb}=2\pi \times 0.5 $MHz, respectively.} 
\label{fig:3}
\end{figure}
 \section{Results and Discussion}
In this section, we present the result of our numerical simulations.
For numerical calculation, we consider the parameters from the recent experiment on a hybrid megnomechanical system~\cite{zhang_cavity_2016, li_phase_2020}:
$\omega_a = 2\pi \times 13.2$GHz, $\omega_b = 2\pi \times 15$MHz, $\kappa_a = 2\pi \times 2.1$MHz, $\kappa_m = 2\pi \times 0.1$MHz, $\gamma_b = 2\pi \times 150$Hz, $D=250~\mu$m, and the magnon-photon coupling is $g_{ma}=2\pi \times 2.0 $MHz.  
In order to study GHS, we consider $\epsilon_{0}=1$, $\epsilon_{1}=2.2$, $d_{1}=4$mm, and $d_{2}= 45 $mm~\cite{zhang_cavity_2016}. We consider the YIG sphere with diameter $D=250$~$\mu m$, spin density $\rho=4.22 \times 10^{27}$~$m^{-3}$ and gyromagnetic ratio $\gamma=2 \pi\times 28$~GHz/T~\cite{zhang_cavity_2016, li_phase_2020}. For these parameters, we choose the drive magnetic field $B_0 \leq 0.5$~mT (which corresponds to $G_{mb}/(2 \pi) \leq 1.5$~MHz) such that the system remains in the stable regime~\cite{lu_exceptional-point-engineered_2021}. 

We first illustrate the output absorption spectrum as a function of effective detuning $x$ in Fig.~\ref{fig:chi}.
The solid curve represents the spectrum in the absence of magnon-phonon coupling ($G_{mb}=0$).
In this condition, only the magnon mode is coupled to the cavity field mode, resulting in the splitting of the output spectrum into two Lorentzian peaks with a single dip at resonance.
This spectrum is known as magnon-induced transparency.
The width of this transparency window depends on magnon-photon coupling $g_{ma}$.
Switching on the magnon-phonon effective coupling to $G_{mb}=2\pi \times 0.5 $MHz by applying an external magnetic field, the single magnon-induced transparency window splits into a double window due to the non-zero magnetostrictive interaction. 
These results are shown by the dashed curve in Fig.~\ref{fig:chi} and known as Magnomechanical Induced Transparency (MMIT).
As the effective coupling strength $G_{mb}$ increases from zero, the height of the central peak starts increasing, along with a slight shift to the left of the resonance point.
Next, we discuss the effects of control field strength, effective cavity detuning, and cavity decay rates, which are responsible for enhanced manipulation
of the GHS at different incidence angles.
For the sake of simplicity, we consider that the incident probe field is a plane wave.

\begin{figure}[t]
\begin{tabular}{@{}cccc@{}}
\includegraphics[width=3.25 in]{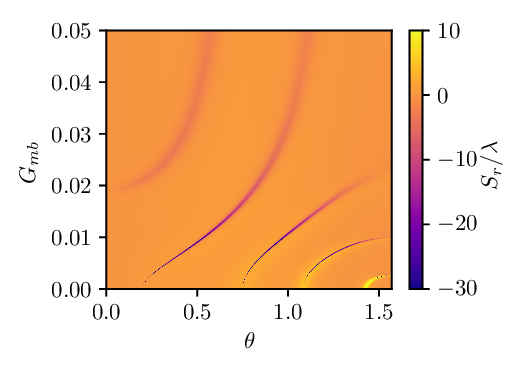}
\end{tabular}
\caption{Contour plot of GHS as a function of incident angle and effective magnomechanical coupling $G_{mb}$ (in the units of $2\pi \times$MHz) at resonance condition $x=0$. Clearly, lower values of magnon-phonon coupling induce larger negative GHS.} 
\label{fig:4}
\end{figure}

\emph{Effects of microwave drive field:}
The strength of the microwave drive field depends on the magnitude of an external magnetic field (\textbf{$E_d \propto B_0$}).
From steady-state dynamical values (See Eq.~(\ref{steady})), it is evident that all excitation modes strongly depend on the strength of the microwave drive field via $m_s$.
Similarly, the effective magnomechanical coupling coefficient $G_{mb}$ is directly proportional to $m_s$ at fixed $g_{mb}$. 
As a result, the output spectrum of the probe field is modified by the strength of the microwave drive field.
We recall that, when $G_{mb}$ is kept zero, the intracavity medium becomes transparent to the probe light beam at effective resonance $x=\delta - \omega_b=0$. 
To see this effect on GHS, we plot the GHS as a function of probe light incident angle $\theta$ (in the units of radian) in Fig.~\ref{fig:3} at resonance condition.
It can be seen from the red curve that the GHS shift is always positive in the absence of microwave driving and exhibits three peaks.
The peak of GHS gets enhanced at a larger incident angle and is maximum at $\theta=1.42$ radian. 

When the effective magnomechanical coupling $G_{mb}$ is turned on through an external microwave drive field, the absorption at resonance $x=0$ starts to appear (See Fig.~\ref{fig:chi}).
The dashed and dotted curves in Fig.~\ref{fig:3} show GHS at $G_{mb}=2\pi \times 0.05 $MHz and $G_{mb}=2\pi \times 0.5 $MHz, respectively.
The other parameters are unchanged.
We note negative GHS in the reflected probe light beam at certain incident angles, $\theta$.  
Switching of GHS from positive to negative is related to the group index of the total cavity system~\cite{wild_goos-hanchen_1982, ziauddin_coherent_2010}.
The group index is defined as the ratio of the speed of light in a vacuum to the group velocity of the reflected field and can be approximated as 
\begin{align}\label{Eg}
 N_{g} \approx \frac{1}{L} \frac{d \phi_r}{d \omega_p},
 \end{align}
where $L=2 d_1 +d_2$ is the total thickness of the cavity. 
The GHS is positive for a positive value of $N_{g}$ and negative for a negative value of $N_{g}$~\cite{ziauddin_coherent_2010}.
It is already established for the case of an atomic medium that the GHS in the reflected beam is negative for absorptive medium and positive for transparent medium~\cite{wild_goos-hanchen_1982}.
Therefore, GHS can be coherently switched from positive to negative via an external microwave drive field. 
The increase of $G_{mb}$ reduces the amplitude of the negative peaks and peaks get shifted to a higher angle as shown in Fig~\ref{fig:3}.
Therefore, in order to understand this effect more clearly, we present the contour plot of GHS as a function of $G_{mb}$ and angle $\theta$ in Fig.~\ref{fig:4}.
The large negative GHS can be obtained at lower values of $G_{mb}$~\cite{note}.
The magnitude of GHS depends on the absorption of the probe field (results shown in Fig.~\ref{fig:chi}) via the reflection coefficient (Eq.~\ref{E3}).
As the value of $G_{mb}$ increases, the probe field gets more absorbed, which results in a decrease in the magnitude of GHS.

\begin{figure}[t]
\begin{tabular}{@{}cccc@{}}
\includegraphics[width=3.250 in]{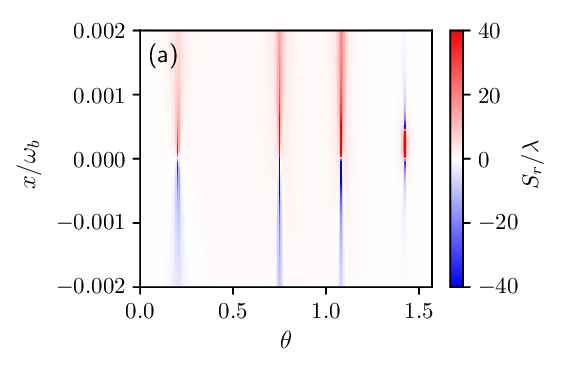}\\
\includegraphics[width=3.250 in]{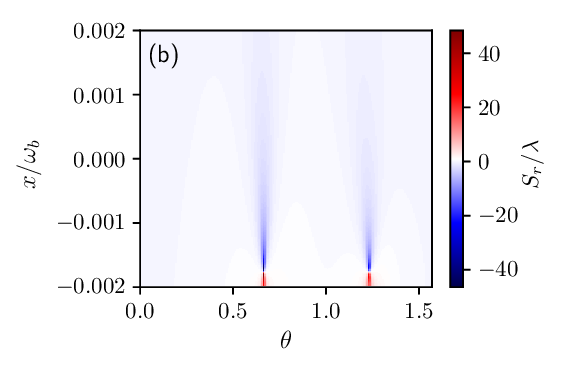}
\end{tabular}
\caption{Contour plot of GHS as a function of incident angle $\theta$ and effective detuning $x$: (a) When $G_{mb}=0$, which indicates that GHS can also be controlled via effective cavity detuning. (b) When $G_{mb}=2 \pi \times 1.0 $MHz, which shows that negative GHS is larger slightly away from the resonance.} 
\label{fig:5}
\end{figure}

\emph{Effects of Detuning:}
Next, we consider the effect of another control parameter, the effective cavity detuning on the reflected GHS.
Figure \ref{fig:5} shows the dependence of the GHS on the effective cavity detuning and incident angles.
We observe the manipulation effect on GHS under different strengths of the microwave control field. 
Fig. \ref{fig:5}(a) shows the results when $G_{mb}$ is kept zero, which means that there is no influence of the microwave drive field on the
cavity. 
At resonance ($x=0$), we observe a sharp transition from positive peaks to negative peaks.  The magnitude of the GHS is higher at detuning relative to the resonance $x=0$ and eventually gets smaller at larger detuning.
These results provide another control mechanism of the GHS from positive to negative by changing the effective detuning in the absence of a microwave drive field.
At a large incident angle, $\theta=1.42$ positive shift appears in a narrow interval of detuning.  
In Fig.~\ref{fig:5}(b), we plot the dependence of the GHS on effective cavity detuning and incident angles of the probe light beam in the presence of the microwave control field.
We note that the transition point from negative to positive GHS peak moves from resonance to negative effective detuning.
As a result, we have negative GHS for a relatively larger range of detuning.
Away from the transition point, towards positive detuning, the amplitude of peaks decreases and peaks get broader.

\begin{figure}[h]
\begin{tabular}{@{}cccc@{}}
\includegraphics[width=3.250 in]{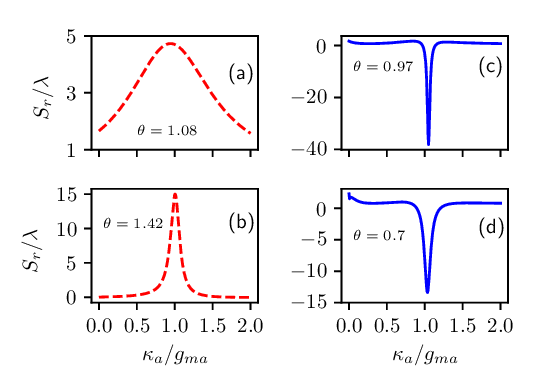}
\end{tabular}
\caption{GHS as a function of cavity decay rate $\kappa_a/g_{ma}$ for (a) $G_{mb}=0$, $\theta=1.08$ 
(b) $G_{mb}=0$, $\theta=1.42$, (c) $G_{mb}=2\pi \times 0.01 $MHz, $\theta=0.97$, and (d) $G_{mb}=2\pi \times 0.015 $MHz, $\theta=0.70$. For these results, we consider the resonant case $x=0$, while the rest of the parameters are the same as given in the text.} 
\label{fig:6}
\end{figure}

\begin{figure}[h]
\begin{tabular}{@{}cccc@{}}
\includegraphics[width=3.250 in]{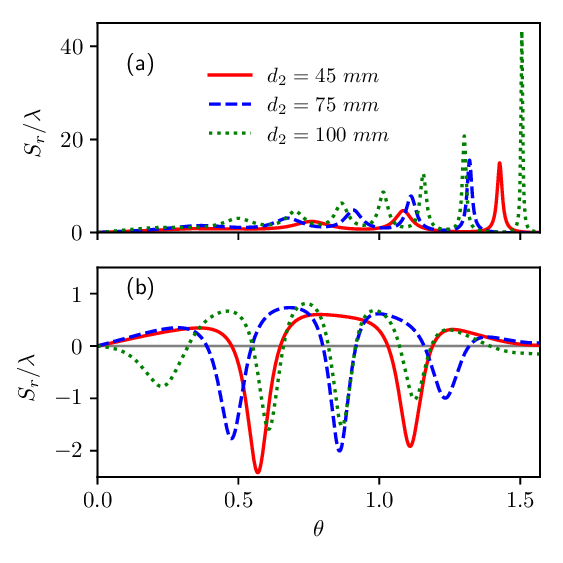}
\end{tabular}
\caption{GHS as a function of incident angle for (a) $G_{mb}=0$ and (b) $G_{mb}=2 \pi \times 0.05$~MHz at three different intracavity medium lengths. Solid, dashed, and dotted curves show GHS at $d_{2}=45~mm$, $d_{2}=70~mm$, and $d_{2}=100~mm$, respectively. Here we consider resonance condition $x=0$. The rest of the parameters are the same as given in the text.} 
\label{fig:7}
\end{figure}

\emph{Effects of weak and strong coupling:}
Indeed, our cavity magnomechanical system is a lossy one because the cavity photons have a limited lifetime after which they decay (lose energy) at the cavity decay rate $\kappa_a$.
The cavity decay rate may be different for different microwave cavities depending on its quality factor $Q$. 
Therefore, it may also affect the reflection coefficient as well as the GHS of the reflected probe light beam. 
Figure \ref{fig:6} shows the behavior of the GHS against the normalized cavity decay rate in units of magnon-phonon coupling $g_{ma}$ at resonance ($x=0$).
The value of ratio $\kappa_a / g_{ma}$ defines the coupling regime of the system.
For instance, strong coupling regime corresponds to $\kappa_a / g_{ma} <1 $ whereas weak coupling regime corresponds to $\kappa_a / g_{ma} > 1$.
Figure \ref{fig:6} (a) and (b) shows the results in the absence of microwave drive when $G_{mb}=0$ at two different angles $\theta=1.08$ and $\theta=1.42$, respectively.
These two angles correspond to the last two peaks of GHS from Fig.~\ref{fig:3} (red curve).
We note that the GHS remains positive for the whole range of $\kappa_a$ considered, here.
However, it peaks around $\kappa_a \approx g_{ma}$ and decreases symmetrically in the weak and strong coupling regime.
This symmetric decrease around $\kappa_a \approx g_{ma}$  becomes very fast, resulting in a narrow spectrum for a larger incident angle (See Fig.~\ref{fig:6}(b)).
Figure \ref{fig:6} (c) and (d) show the results in the presence of microwave drive when $G_{mb}=2\pi \times 0.01 $MHz and $G_{mb}=2\pi \times 0.015 $MHz, respectively. 
We choose the maximum literal shift angle $\theta=0.97$ and $\theta=0.70$ from the results of Fig.~\ref{fig:4}.
The GHS shift is negative and has a dip around $\kappa_a \approx g_{ma}$. 
Again, the width of the GHS spectrum is narrow at a larger angle.
At a fixed incident angle, the peak value of GHS depends on the reflection coefficient $R$ as evident from Eq.~\ref{E2}.
At $\kappa_a \approx g_{ma}$, $R$ approximately approaches zero, which results in the peak value of GHS.

So far, we have discussed the effect of the drive field, which is directly related to the strength of effective magnomechanical coupling coefficient, $G_{mb} = g_{mb}m_s$ via $m_s$ (See Eq.~(\ref{steady})).
Another important parameter worth studying is the cavity size.
There is indeed a strong dependence of GHS on the cavity structure, such as the total thickness of the cavity $L=2d_1+d_2$.
Therefore, we plot the GHS as a function of incident angle $\theta$ at three different intracavity medium lengths $d_2$ for $G_{mb}=0$ in Fig~\ref{fig:7}(a) and $G_{mb}=2 \pi \times 0.05$MHz in Fig~\ref{fig:7}(b) at $x=0$.
Both figures show that an increase in cavity length introduces more resonance peaks/dips in the GHS as a function of incidence angle.
This dependence of GHS on medium thickness in cavity magnomechanics is analogous to the behavior previously discussed in GHS studies in the atomic medium~\cite{ziauddin_coherent_2010}.
\section{Conclusion}
In summary, we have theoretically investigated the GHS in a cavity magnomechanical system where the magnon mode is excited by a coherent microwave control field.
We noted the coherent manipulation of the GHS by the control microwave field. 
The GHS is positive at resonance in the absence of a control field.
By turning on the control field, we show that GHS changes from positive to negative.
Similarly, by modifying the effective cavity detuning in the absence of a control field, we have shown the behavior of the GHS changing from positive to negative.
For instance, positive detuning gives positive GHS while negative detuning gives negative GHS.
This symmetric behavior of GHS around resonance, however, can be changed by turning the control field on.
We also identify the optimum ratio of microwave photon lifetime to magnon-photon coupling to maximize the GHS.
It is shown that the larger incident angles are more sensitive to this optimal ratio as compared to smaller angles.

In this work, we have considered a generic cavity magnomechanical system to demonstrate the idea of GHS.
We mainly used magnomechanical induced transparency configuration to observe the changes in the GHS of the reflected probe field.
The connection between the two concepts of magnomechanical induced transparency (MMIT) and magnomechanical Goos-Hanchen shift is qualitatively the same as the connection between electromagnetically induced transparency (EIT) in atomic media and the corresponding Goos-Hanchen shift of the probe field due to the atomic media~\cite{wang_control_2008, ziauddin_coherent_2010}.
It is important to note that the idea of tuneable magnomechanical induced transparency spectrum has been demonstrated in recent experiments~~\cite{zhang_cavity_2016, Zhao-PhysRevApplied}.
The experimental realization of GHS in cavity magnomechanics may require design modifications in the typical experimental systems or consideration of other hybrid cavity-magnomechanical systems~\cite{ZARERAMESHTI20221, Fan_2023, Fan-PRA2023, Hatanaka_2023, Yang_2024, Zhu-20-optica}.
We therefore believe that our analysis may be useful to investigate GHS in cavity magnomechanical systems and may potentially lead to the development of microwave devices using GHS.
Some possible examples could be quantum switching and microwave high-precision measurement sensing.

\section*{ACKNOWLEDGEMENTS}
We acknowledge the fruitful discussions with Dr. Muzammil Shah and Dr. Muhib Ullah.

\bibliographystyle{apsrev4-2}
\bibliography{Ref.bib}

\begin{thebibliography}{62}
\bibitem [{not({\natexlab{b}})}]{note}%
  \BibitemOpen
  \href@noop {} {\bibinfo {title} {Because {GHS} peaks become very sharp at certain values of parameters. {W}e used very small step size (of both variables involved) to get smooth and high resolution plots in figure (4) and (5). {A}s a result, some peaks/dips becomes quite sharp with large amplitudes. {W}e therefore truncated the color bar limit to demonstrate the effect clearly.}} ({\natexlab{b}})\BibitemShut {NoStop}%
\end{thebibliography}

\end{document}